Multi-ferroelectricity in charge ordered $LuFe_2O_4(LuFeO_3)_n$ with n=1

H. X. Yang [1], Y. Zhang[1], C. Ma[1], H. F. Tian[2], Y. B.Qin[1], Y. G. Zhao[2] & J. Q. Li †

[1] Beijing National Laboratory for Condensed Matter Physics, Institute of Physics, Chinese Academy of Sciences, Beijing 100080, P. R. China.

[2] Department of Physics, Tsinghua University, Beijing 100084, China

† To whom correspondence should be addressed. E-mail: ljq@aphy.iphy.ac.cn

**Abstract**

Electronic ferroelectricity from charge ordering (CO) is currently a significant issue that has been extensively investigated in the charge/spin frustrated $LuFe_2O_4$ system. Chemical substitution and structural layer intercalation have been considered as potentially effective approaches in our recent study, and we herein report on the structural and multi-ferromagnetic properties in the layered series of $LuFe_2O_4(LuFeO_3)_n$ (n=0, 1, and 2). Experimental investigations reveal that the ($RFeO_3$)-layer intercalation results in notable changes in ferroelectricity, magnetic properties and CO features. These results demonstrate that further exploration of this series of layered members might prove fruitful.

Multi-ferroelectric materials have stimulated considerable interest because of remarkable properties that could yield immense benefits in advanced materials for technical applications in modern electronic devices such as memory elements, filtering and switching devices[1]. Electronic ferroelectricity from charge ordering (CO) is currently a significant issue that has been extensively investigated in the charge/spin frustrated $LuFe_2O_4$ system; moreover, a giant magneto-dielectric response was observed in this material at room temperature[2-6]. Recently, experimental measurements on $(Pr,Ca)_3Mn_2O_7$ demonstrated that an electrically polarized state can exist in the charge/orbital ordering phase, exhibiting certain observable anomalous physical properties at critical temperatures[7]. But for technology applications to be viable, great improvements are needed in both sensitivity and stability of these new materials. Chemical substitution and structural layer intercalation have been considered as potentially effective approaches in our recent study, and we herein report on the structural and multi-ferromagnetic properties in the layered series of $LuFe_2O_4(LuFeO_3)_n$ (n=0, 1, and 2). Experimental investigations reveal that the $(RFeO_3)$-layer intercalation results in notable changes in ferroelectricity, magnetic properties and CO features. These results demonstrate that further exploration of this series of layered members might prove fruitful.

Structurally, each layered $LuFe_2O_4(LuFeO_3)_n$ phase is simply constructed by an alternate stacking of $LuFe_2O_4$ and $(LuFeO_3)_n$ blocks along the c-axis direction[8]. Schematic structures of the layered family are shown in Fig.1 (a) for n=0 and n=1. It can be seen that all n=0, 1 and 2 phases contain the charge frustrated $Fe_2O_{2.5}$ layers (W-layer), while the

n=1 and 2 phases also contain the $FeO_{1.5}$ layers (V-layer). The equal numbers of $Fe^{2+}$ and $Fe^{3+}$ in a W-layer can be ordered and result in electronic ferroelectricity as extensively discussed in the context of the $LuFe_2O_4$ (n=0) phase[9]. The valence states of Fe ions in the $Lu_2Fe_3O_7$ are much more complex. (The average valence state for Fe is estimated to be about 2.67 for n=1.) The measurements of the Mössbauer spectrum suggested that the Fe ions in the V layer have valence state of $Fe^{3+}$ and therefore CO probably occurs in the W layers[10]. However, our recent study on the n=1 phase revealed that the charge disproportionation also appears in the V layers recognizable as clear CO modulations, suggesting that the intercalation of $LuFeO_3$ blocks could yield notable effects on the interaction among layers and cause further charge redistribution on the Fe sites. The intercalated $(RFeO_3)n$ blocks, as a tunable chemical/structural factor, can reduce or increase the electrons in the charge frustration layers (the exact value is in question). Our recent study also shows that it is difficult to produce a single-phase sample of the n=2 compound [11]. A mixture of phases is usually seen, each having its own number of Fe-O planes and Lu-O planes per unit cell. This syntactic intergrowth has made it difficult to do good physics experiments. Hence, in the present paper, we will mainly report on our experimental results on the $Lu_2Fe_3O_7$ (n=1) phase.

Polycrystalline samples of $Lu_2Fe_3O_7$ materials were synthesized from stoichiometric mixtures of $Lu_2O_3$ (99.99%) and $Fe_2O_3$ (99.99%). After mixing and milling, the raw materials were sintered by conventional solid-state reaction under a controlled oxygen partial pressure atmosphere using a $CO_2$-$H_2$ mixture at 1473K for 48h. The XRD measurements indicate the diffraction pattern taken from the $Lu_2Fe_3O_7$ samples can be confidently indexed as a hexagonal cell with a P6$_3$/mmc space group and lattice parameters of a =3.455 Å and c = 28.444Å, no peaks from impurity phases are observed. Transmission electron microscopy (TEM) investigations were performed on a Tecnai F20 (200kV) electron microscope equipped with a high-temperature holder. Figure 1(b) shows an HRTEM image taken along the [100] zone axis direction, clearly displaying the atomic structural features of the Lu$_2$Fe$_3$O$_7$ crystal. This image was obtained from a thin region of crystal under the defocus value at around the –40 nm. Image simulations suggest that the metal atom positions are recognizable as white dots; the Lu and Fe positions are clearly illustrated in the experimental image.

Figure 2 shows the zero-field cooling (ZFC) and field cooling (FC) magnetizations of Lu$_2$Fe$_3$O$_7$ as a function of temperature at an applied field of 100 Oe. Results obtained from a LuFe$_2$O$_4$ sample are also displayed for comparison. Despite the recognizable differences in the critical temperatures between the n=0 and 1 phases, these phases have markedly similar magnetic features. e.g. a clear cusp-like peak appears at around 255K for Lu$_2$Fe$_3$O$_7$ in the ZFC data and at 234K for LuFe$_2$O$_4$. In previous studies, this type of temperature dependence of magnetization was respectively discussed in several temperature ranges [13]. These facts directly suggest that the spin frustration resulting in complex magnetic features, as typically discussed in the n=0 phase, also play a critical role in the present system.

Figure 3 (a) shows the temperature dependence of the real parts of the dielectric constant (ε') of Lu$_2$Fe$_3$O$_7$, illustrating the AC dielectric dispersions for different frequencies. Dielectric dispersion varies dramatically between 90K and 110K depending on frequency. This ferroelectric data is comparable with results obtained from $LuFe_2O_4$[4], but the temperature (and also frequency) dependence of the dielectric data has a notable difference. We therefore conclude that the charge frustration and CO occurring at low temperatures are affected by the $LuFeO_3$-block intercalation, as is also demonstrated by our in-situ TEM investigations.

Figure 3(b) shows the data of the loss factor tan δ in the low temperature range. The loss peak position within 95K to 110K shifts to higher temperatures with the frequency increase. Careful analysis reveals that this phenomenon can be well described by the Arrhenius formula $f = f_0 \exp(-Ea/KT)$, where $f_0$ is the pre-exponential term, $Ea$ is the activation energy, and $K$ is the Boltzmann constant. The fitting parameters for our experimental results were estimated to be $E_a$=0.179eV and $f_0$=5.8×10$^{13}$ Hz. The activation energy $E_a$ is somewhat smaller than that for the $LuFe_2O_4$ ferroelectricity system (0.29eV)[14].

In order to well-characterize the ferroelectric properties of the n=1 phase, we performed a series of measurements on the ferroelectric *P-E* loops at different temperatures. As a result, the P-E curves at room temperature show visible instability due to the dynamic fluctuation of local polarizations and the relatively high conductivity, and well-defined P-E ferroelectric hysteresis loops can be obtained at low temperature. Figure 4a presents a P-E response at a temperature of 150K from a well-characterized $Lu_2Fe_3O_7$ sample. It is recognizable that this hysteresis loop exhibits visible distinctions from an ideal square-type response owing to the conducting effects as similarly observed in $LuFe_2O_4$[15]. Figure 4b shows the magnetization curve at a temperature of about 80K. The magnetization features a minor hysteresis loop indicating ferromagnetic properties at low temperatures. Magnetization increased with the H field, but did not saturate at very high fields, suggesting a complex magnetic nature in this kind of materials. In present case, we can estimate the spontaneous magnetization *Ms* by extrapolating the linear portion of the magnetization curve to *H*=0. The estimated value of *Ms* is about 0.5emu/g.

The CO states and their correlation with the ferroelectricity are the key issues for understanding the significant properties observed in this electronic ferroelectric system, as extensively discussed in the $LuFe_2O_4$ material[9]. It is clearly demonstrated that the complex super-lattice reflections arising from ordered arrangements of the mixed valence Fe ions provides direct structural evidence for characterizing the ferroelectric polarization[9]. We therefore have carried out a series of in-situ TEM investigations focused on the super-lattice reflections in $Lu_2Fe_3O_7$. The results demonstrate that the CO phenomena can be observed in both $FeO_{1.5}$ and the charge frustrated $Fe_2O_{2.5}$ layer.

The better and clearer view of the CO modulations were obtained along [1$\bar{1}$0] zone-axis direction as shown in Fig. 5a, demonstrating the presence of clear superstructure spots following the main diffraction spots. The main spots with relatively strong intensity in the diffraction pattern can be indexed very well on the hexagonal unit cell with lattice parameters of a =3.455Å and c = 28.444Å (space group of $P6_3/mmc$). In comparison with the data as discussed for the $LuFe_2O_4$ phase[9], notable differences of the superstructure reflections are clearly recognizable in both direction and periodicity, directly suggesting that the CO behavior in the n=1 phase is notably influenced by the $LuFeO_3$-block intercalation. Careful investigations suggest that the superstructure spots can be assigned respectively to two incommensurate modulations ($q_1$ and $q_2$) as clearly illustrated in the enlarged pattern of Fig. 5b. The modulations have wave vectors of $q_1$=(1/3, 1/3, 1) + δ and $q_2$= (1/3, 1/3, 0) - δ, where δ is the incommensurate parameter with a value of about δ =1/10. These modulations should correspond with certain kinds of $Fe^{2+}$ and $Fe^{3+}$ orders appearing in both the charge frustrated $Fe_2O_{2.5}$ and the $FeO_{1.5}$ layers as discussed in the following context.

In order to characterize the origins of $q_1$ and $q_2$ modulations, a series of high-resolution TEM images have been taken to reveal the structural features corresponding with these modulations. Fig. 5c shows a high resolution TEM taken with a large defocus value of about –60nm, which therefore clearly shows up the double $Fe_2O_{2.5}$ and the single $FeO_{1.5}$ sheets as white layers. It is easily recognizable that structural modulations appear in both the charge frustrated double $Fe_2O_{2.5}$ layers and the $FeO_{1.5}$ layers. We later performed Fourier transformations on the experimental images to obtain the filtered images for the $q_1$ and $q_2$ modulation, respectively. The results suggest that structural features of $q_2$ appear in all atomic (Lu/Fe) layers, while on the other hand, the $q_1$ modulation is chiefly visible in the charge frustrated $Fe_2O_{2.5}$ layers.

Though we have made numerous attempts to interpret these two modulations accompanying $Fe^{3+}$ and $Fe^{2+}$ ordering, we found that it is really difficult to concurrently interpret these two modulations by a simple structural model, hence, we first focus our attention on the CO states in the frustrated layers corresponding with the $q_1$= (1/3, 1/3, 1) + δ. Based on our previous study for the n=0 phase, the $q_1$ can be reasonably interpreted by an ionic order of $Fe^{2+}Fe^{2.5+}Fe^{3+}$ within the double $Fe_2O_{2.5}$ layers. Fig. 5d shows a schematic CO pattern in accordance with our experimental observations. It is remarkable that positive charges ($Fe^{3+}$ sites) and negative charges ($Fe^{2+}$ sites) are crystallized in parallel charge stripes. This CO pattern shows up as a clear monoclinic feature with an evident electric polarity. It is noted that the typical structure of $Lu_2Fe_3O_7$ is a hexagonal symmetry with the $P6_3/mmc$ space group, and therefore, the $q_1$ modulation, together with the local electric polarization, could appear evenly in three crystallographically equivalent <113>-directions around the **c-** axis, and the resultant spontaneous electric polarization would be aligned with the c-axis as similarly discussed for $LuFe_2O_4$[9]. Of course, the real CO pattern and systematic ferroelectric polarization in the n=1 phase also correlate with the $q_2$ modulation. A further study to reveal the correlation between $q_2$ and electric polarization is still in progress.

The critical temperatures for the CO transitions, as well as ferroelectric phase transitions, are also significant issues for $Lu_2Fe_3O_7$ and addressed by the present study. We thus carried out an in-situ heated TEM observation to reveal the temperature dependence of the superstructure reflections. The experimental results, as shown in the supplementary information, demonstrate that the intensities of the superstructure reflections decrease progressively with increasing temperature, and typically disappear at a critical temperature of about 730K, which is much higher than the critical temperature of $T_{co} \approx 530K$ in $LuFe_2O_4$. TEM investigations also reveal that this transition is reversible. Though remarkable hysteresis often exists, the superstructure reflections become visible again as the temperature is lowered below the critical temperature.

In summary, the $LuFe_2O_4(LuFeO_3)_n$ with n=1 shows remarkable multi-ferroelectricity in correlation with the essential CO nature existing in this kind of materials. Our systematic measurements demonstrate that the intercalation of the $LuFeO_3$-block, as a chemical/structural tunable factor, results in notable effects on the electronic ferroelectricity and also causes charge redistribution on the Fe sites. As a result, the charge ordered state in the n=1 phase – characterized by two incommensurate modulations, $\mathbf{q_1}$ and $\mathbf{q_2}$ – is much more complex than that in the n=0 phase. In contrast with the diffuse satellite streaks in the n=0 phase at room temperature, the n=1 phase often gives rise to sharp superlattice spots, suggesting a longer CO coherence length, which would favor the existence of larger ferroelectric domains in the present material. Hence, it is expected that further control of the $LuFeO_3$ block number n≥2 in this layered family should help toward achieving useful multi-ferroelectric characteristics at the desired operating conditions of room temperature.

**Acknowledgments** We would like to thank Dr. J. Yu and Dr. R. Walton for fruitful discussions. The work reported here is supported by the National Natural Science Foundation of China, the Chinese Academy of Sciences and the Ministry of Science and Technology of China.

**Author Information** The authors declare no competing financial interests. Correspondence and requests for materials should be addressed to J.Q. Li (ljq@aphy.iphy.ac.cn).

Supplementary information accompanies this paper.

Figure captions

**Figure 1** (a) Structural models schematically illustrating the Lu-O layer, Fe-O double and single layer stacking alternatively along the c-axis for $LuFe_2O_4(LuFeO_3)_n$ phase ( n=0 and n=1); O-atoms are omitted for clarity. (b) [100] zone-axis HRTEM image clearly displaying the atomic structure of $Lu_2Fe_3O_7$.

**Figure 2** The zero-field cooling (ZFC) and field cooling (FC) magnetizations of (a) $Lu_2Fe_3O_7$ and (b) $LuFe_2O_4$ as a function of temperature at an applied field of 100Oe. Changes of magnetic properties are demonstrated.

**Figure 3** Temperature dependence of (a) dielectric constant and (b) dielectric loss for $Lu_2Fe_3O_7$ ceramics at different frequencies.

**Figure 4** (a) The Polarization versus electric field hysteresis loop collected at a frequency of 100Hz at 150K. (b) The magnetization curve at a temperature of about 80K with a minor hysteresis loop indicating ferromagnetic properties in low temperatures.

**Figure 5** Electron diffraction patterns of $Lu_2Fe_3O_7$ taken along [1$\bar{1}$0] zone-axis directions at room temperature, showing clear superstructure spots. (b) The better and clearer view of the CO order modulations. (c) [1$\bar{1}$0] zone-axis HRTEM image showing the CO states. (d) A structural model for the ionic order of $Fe^{2+}$ and $Fe^{3+}$.

**Figure** 1

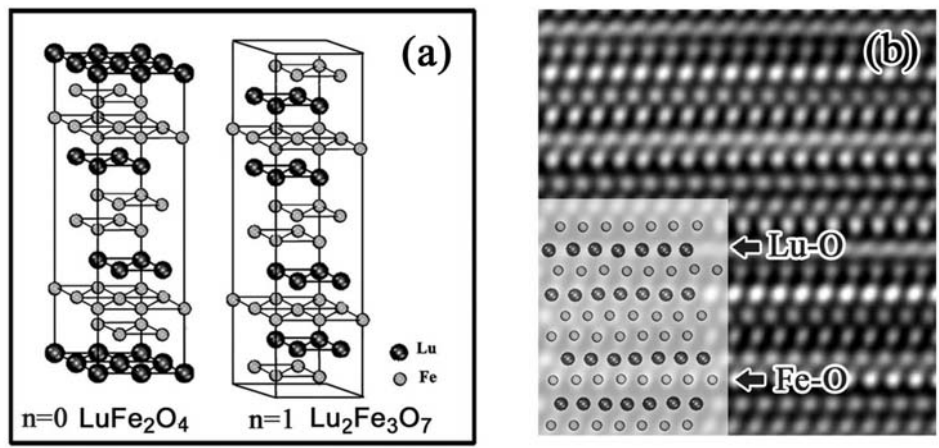

**Figure 2**

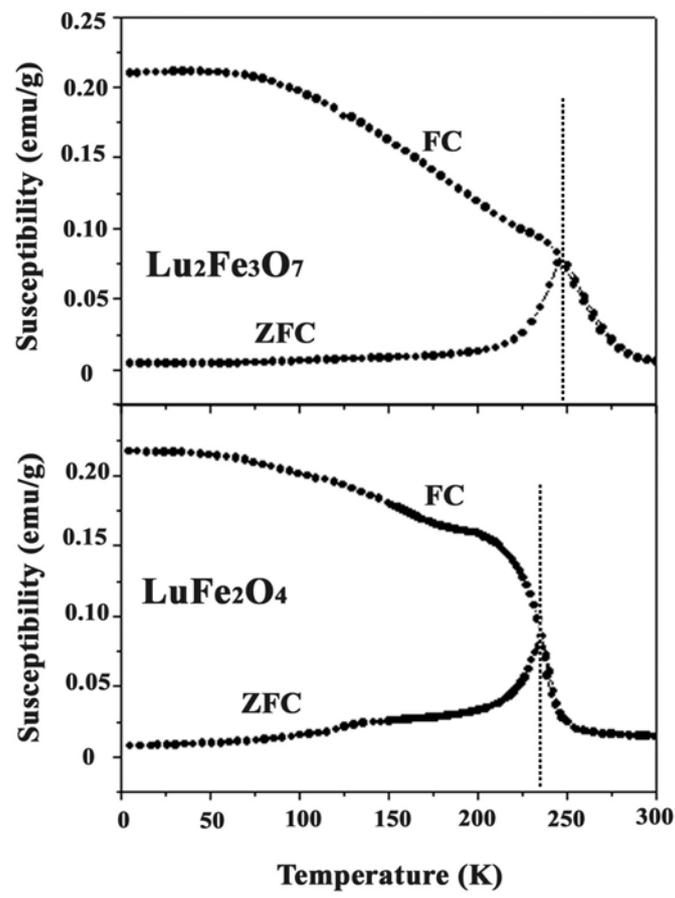

**Figure 3**

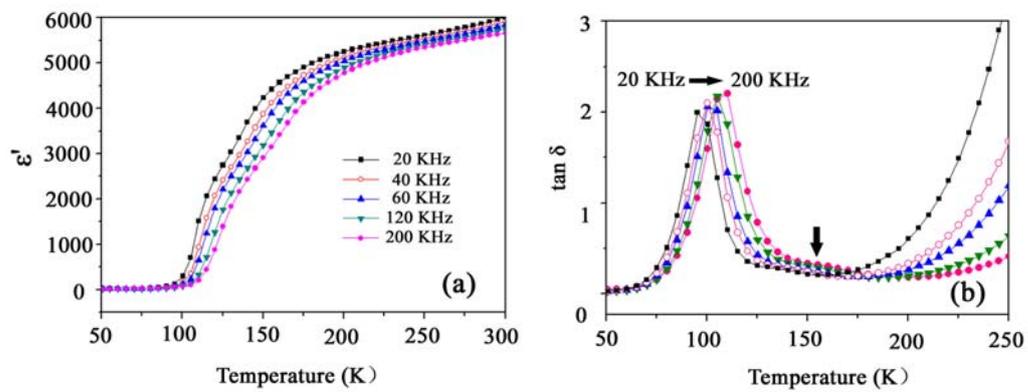

**Figure 4**

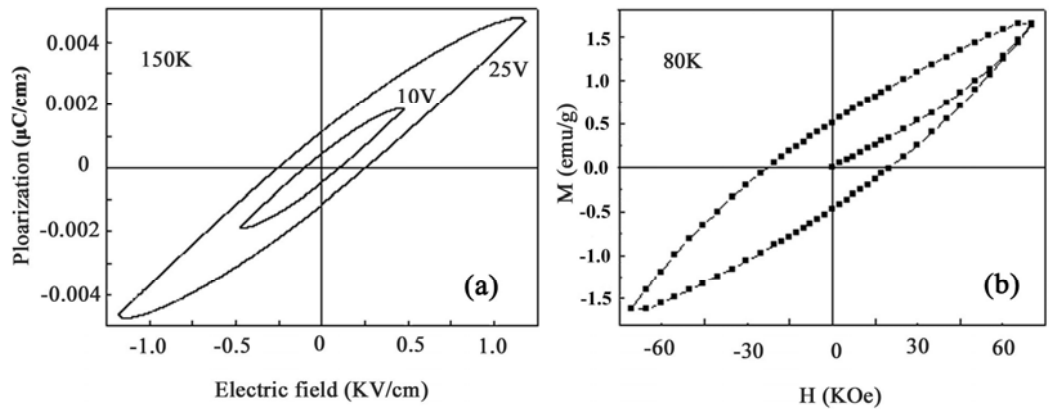

**Figure 5**

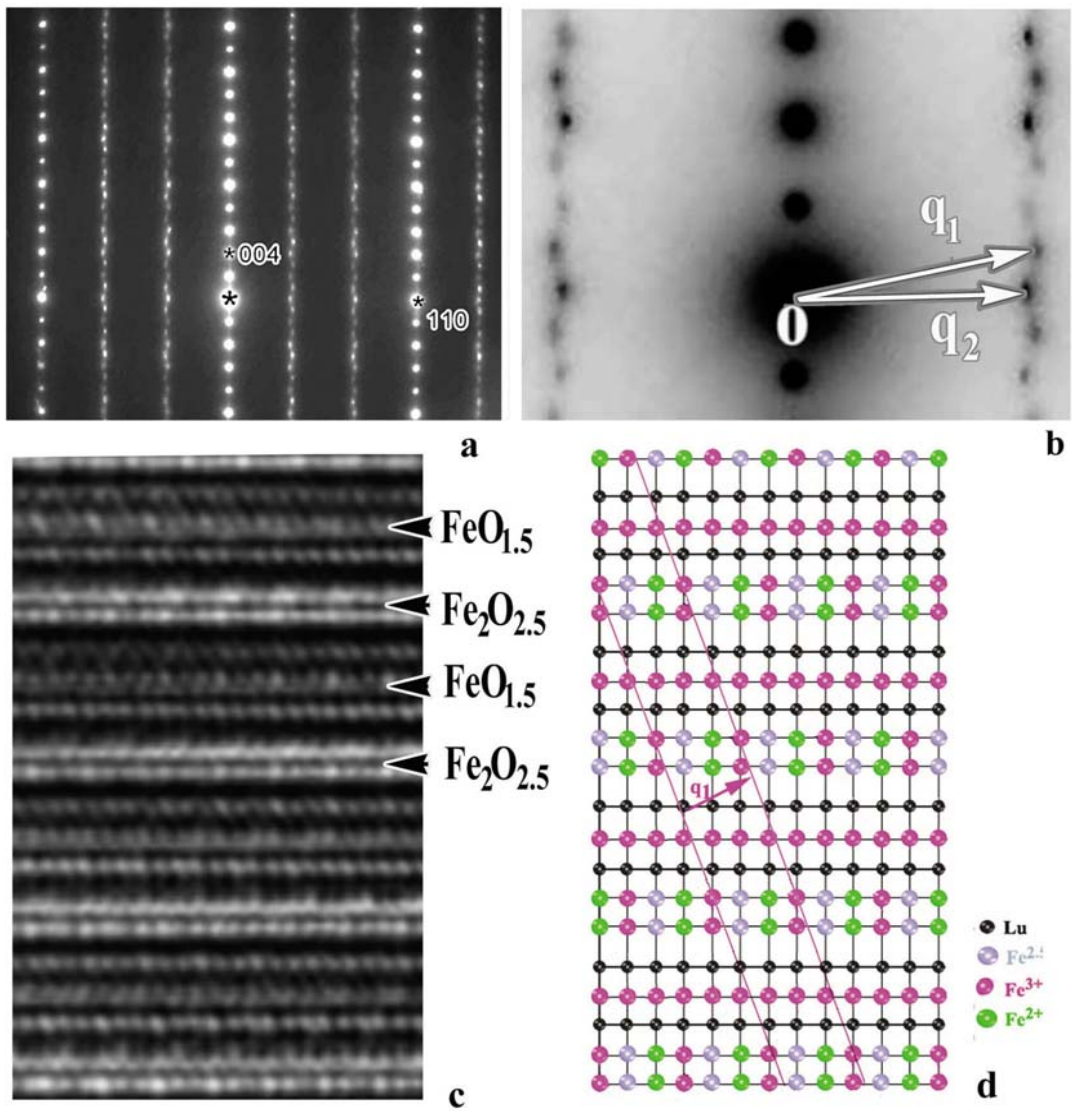

**Supplementary information**

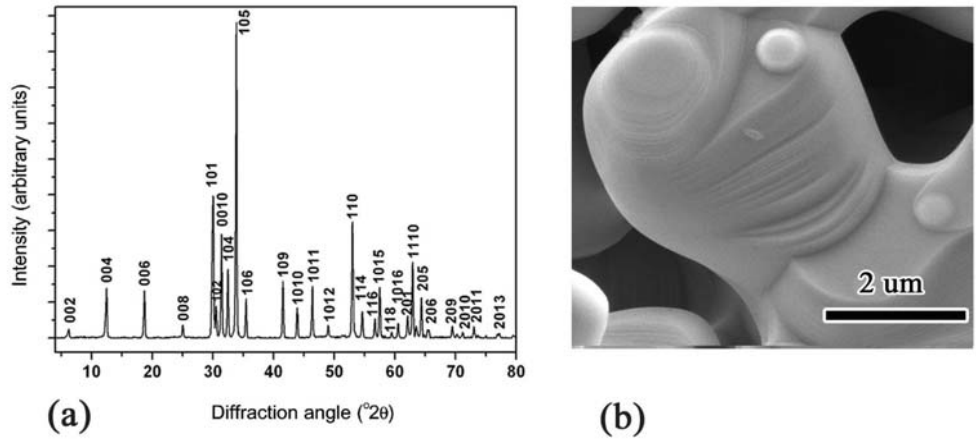

FIG. 1.  (a) X-ray diffraction pattern from a $Lu_2Fe_3O_7$ sample. (b) An SEM image of $Lu_2Fe_3O_7$, clearly illustrating layered structural features in this material.

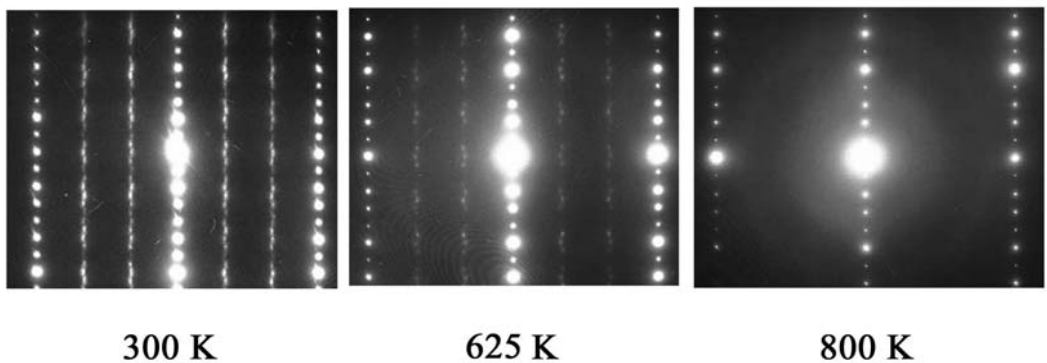

FIG. 2. The temperature dependence of the superstructure reflections from 300K to 800 K. The intensities of superlattice reflections become weaker and more diffuse with increasing temperature, and disappear above 730K.